\documentclass[referee]{raa}
\usepackage{graphicx,times,amsmath,amssymb,natbib}             
\usepackage{natbib}
\bibpunct{(}{)}{;}{a}{}{,}
\usepackage{color}
\usepackage{subfigure}
\usepackage{upgreek}
\usepackage[a4paper=true,dvipdfm=true,pagebackref=true]{hyperref}
\hypersetup{pdftitle = The title of my PDF, pdfauthor = My name, pdfsubject= The subject, pdfkeywords = keyword1 keyword2 keyword3} 
\hypersetup{colorlinks = true, linkcolor = green, anchorcolor = red, citecolor = blue, filecolor = red, pagecolor = red, urlcolor = red}
\begin{document}

\title{Log-normal flux distribution of bright \emph{Fermi} blazars}
	\volnopage{Vol.0 (200x) No.0, 000--000}      
	\setcounter{page}{1}          
	
    \author{Z. Shah\inst{1} \and N. Mankuzhiyil\inst{2} \and  A. Sinha\inst{3,4} \and R. Misra\inst{3} \and S. Sahayanathan\inst{2} \and N. Iqbal\inst{1}
    } 
	\institute{ Department Of Physics, University of Kashmir, Srinagar-190006, India; {\it shahzahir4@gmail.com}\\
	\and
	    Astrophysical Sciences Division, Bhabha Atomic Research Centre, Mumbai-400085, India; {\it nijil@barc.gov.in}\\
	\and
	   Inter-University Center for Astronomy and Astrophysics, Pune-411007, India\\
	\and
	  now at: AstroParticule et Cosmoligie, CNRS/University Paris Diderot, 10 Rue Alice Domon et Leonie Duquet, Paris 75013\\
}

\abstract{
	We present the results of the $\gamma$-ray flux distribution study on the brightest blazars which are observed by the \emph{Fermi}-LAT. We selected 50 brightest blazars based on the maximum number of detection reported in the LAT third AGN catalog. We performed standard unbinned maximum likelihood analysis on the LAT data during the period between August 2008 and December 2016, in order to obtain the average monthly flux. After quality cuts, blazars for which at least 90$\%$ of the total flux was survived were selected for the further study, and this includes 19 FSRQs and 19 BL Lacs. The Anderson-Darling and $\chi^2$ tests suggest that the integrated monthly flux follow a log-normal distribution for all sources, except for three FSRQs for which neither a normal nor a log-normal distribution was preferred. A double log-normal flux distribution tendency were observed in these sources, though it has to be confirmed with improved statistics. 
	We also found that, the standard deviation of the log-normal flux distribution increases with the mean spectral index of the blazar, and can be fitted with a line of slope $0.24\pm0.04$.
	We repeat our study on three additional brightest unclassified blazars to identify their flux distribution properties. Based on the features of their log-normal flux distribution, we infer these unclassified blazars may be closely associated with FSRQs. We also highlight that considering the log-normal behavior of the flux distribution of blazars, averaging their long term flux in linear scale can largely under estimate the nominal flux and this discrepancy can propagate down to the estimation of source parameters through spectral modeling.
}
\authorrunning{Z. Shah et al}
\titlerunning{Blazar $\gamma$-ray flux distribution}

 \maketitle
\section{Introduction}  
Blazars are subclass of Active Galactic Nuclei (AGNs) with their relativistic jets pointing towards line of sight of the observer \citep{Blandford1979}. Even though the mechanism behind the formation of relativistic jets is  not fully understood yet,  it is most likely related to the focusing properties of the fully ionized, rotating accretion disk \citep{Blandford1977}. 
Blazars  include  BL Lac objects and Flat-Spectrum  Radio  Quasars  (FSRQs), where the significant  difference  between  the two classes being their optical  emission/absorption lines, which are strong for FSRQs, while weak or absent for BL Lacs \citep{Urry1995}.  

The spectral energy distribution (SED) of blazars consists of two broad emission components, where the low energy component peaks at optical to X-ray band, while the high energy component peaks at MeV to TeV band. BL\,Lac objects are further subdivided based on the peak frequency ($\nu_s$) of their low energy component namely, high energy peaked BL\,Lac (HBL; $\nu_s > 10^{15.3}$ Hz), intermediate energy peaked BL\,Lac (IBL; $10^{14}<\nu_s\leq10^{15.3}$ Hz), and low energy  peaked BL\,Lac (LBL; $\rm \nu_s\leq10^{14}$ Hz) \citep{Fan2016} . In case of FSRQs, $\nu_s$ usually falls at relatively lower frequencies ($\lesssim10^{14}$ Hz).
The low energy component of the blazar SED is commonly attributed to the synchrotron emission due to the interaction of relativistic electrons in the jet magnetic
field; whereas the high energy component is explained as inverse Compton (IC) scattering process.  If the target low energy photons for the IC process is the
synchrotron photon itself then the IC mechanism is called  Synchrotron Self Compton (SSC; \citealt{Marscher1985,Band1985}).  On the other hand, if the  photon 
origin is external to the jet, e.g. broad line region (BLR), obscuring torus, Cosmic Microwave Background (CMB) etc., then the process is called external Compton (EC)
mechanism \citep{Dermer1992,Sikora1994,Shah2017}). Alternate to this leptonic interpretation of the high energy emission, hadronic models involving nuclear cascades
were also put forth and are successful in explaining many observed features of blazars \citep{Mannheim1992,Bottcher2007}. 

One of the distinct property of blazars is their rapid flux and spectral variability across the entire electromagnetic spectrum on time 
scales ranging from minutes to years.  Though the cause of variability is still not well understood, plausible clues can be obtained by studying 
the long term flux distribution of blazars.
Such studies have been performed in detail at X-ray energies for Seyfert galaxies and X-ray binaries, where the emission at these energies is 
dominated by the accretion disk or its corona.
The X-ray flux of  Seyfert\,1 IRAS\,13224-3809 using ASCA observations in different epochs, exhibit a log-normal distribution \citep{Gaskell2004}. In another 
study, \cite{Uttley2005} found that the X-ray flux of Seyfert\,1   NGC\,4051 also shows a log-normal distribution, which was comparable to the X-ray flux of 
the black-hole X-ray binary Cyg\,X-1. Linear relationship between the optical flux and the corresponding variation were noticed in Seyfert\,1 NGC\,4151
\citep{Lyutyi1987}, which in turn is an indication of log-normality of flux distribution.  
A similar relationship was also noticed in X-ray band in both  Seyfert\,1  Mrk\,766 \citep{Vaughan2003a}, and  Seyfert\,2 MCG\,6-30-15 \citep{Vaughan2003b}.
The log-normality of flux distribution in a blazar  was first detected in   BL Lacertae, from the RXTE observations \citep{Giebels2009}.  This result is
particularly interesting since for blazars, X-ray emission originate from jets rather than the accretion disk or its environment. Hence, this result
may hint the plausible disk-jet connection in blazars, which is still not clearly understood. The log-normality was later observed in many blazars at different
energies. For instance, such behavior was inferred in Mrk\,421 and Mrk\,501  at Very High Energy (VHE $>$100\,GeV) band, though the data was
noncontinuous \citep{Tluczykont2010}. Similarly, the 4-year flux distribution of blazars given in the third Fermi-LAT catalog of AGNs (3LAC; \cite{fermi3agn}), 
showed a log-normal behaviour.
While quantifying the flux variability in  Mrk\,421, \cite{Sinha2016} also noticed a log-normal flux distribution (more than normal) trend, through out the frequencies from radio to VHE. 
On the contrary, a detailed multi-wavelength study of FSRQ PKS\,1510-089,  \cite{Kushwaha2016} found that the flux 
distribution follow two distinctive log-normal profiles in both optical and $\gamma$-rays, while X-ray flux distribution follow a single log-normal distribution. 
Interestingly, the $\gamma$-ray flux distribution of the same source, obtained from a near continuous data during August\,2008-October\,2015, was well fitted by
a log-normal distribution and  similar was the case of HBL Mrk\,421 and FSRQs B2\,1520+31. On the other hand, the $\gamma$-ray flux distribution 
of FR\,I radio galaxy NGC\,1275  was not able to be represented by a log-normal or normal function, even though the rms increases linearly with flux \citep{Kushwaha2017}.

In this work, we aim to study the flux distribution properties of the brightest Fermi blazars using the data collected in more than 8 years. We also investigate the associated spectral properties of these brightest blazars. Further, we examine the above properties in order to associate the unclassified blazar types (BCUs) with the known blazar classes. We select  bright blazars from the 3LAC, and analyze the  data (described in Sect.\,2). In order to overcome the effect of short-term flux variations, which are most likely associated with the change in the emission region geometry, we consider the flux in monthly bins for our study.  After analyzing the features of the flux distribution, and verifying the log-normality (Sect.\,3),  we study the association of flux distribution with spectral properties (Sect.\,4). The results and possible implications are discussed in Sect.\,5.

\section{\emph{Fermi}-LAT analysis} 
The Large Area Telescope (LAT) on board \emph{Fermi} satellite is a pair conversion detector \citep{Atwood2009} with an effective area $\sim 8000 \,\rm cm^2/GeV$ photon, and field of view $\sim 2.4 $sr, in the energy range from 20\,MeV to more than 300\,GeV, which scans entire sky in every 3 hours. We made a primary selection of 25  FSRQs and 25 BL\,Lacs from the four year NASA'S Fermi 3LAC interactive table \footnote{https://fermi.gsfc.nasa.gov/ssc/data/access/lat/4yr$_-$catalog/3FGL-table/}. The selection was based on the criteria such that the chosen FSRQs and BL\,Lacs should have monthly averaged photon flux $\rm > 6.5\times 10^{-9}\,photons\,cm^{-2} s^{-1}$ and $\rm > 5.5\times 10^{-9}\,photons\,cm^{-2} s^{-1}$ respectively, and the number of upper limits (i.e, non detections) should be less than or equal to 4. We have then downloaded the first 8.4 years of data ( from 2008 August to 2016 December) for the selected sources. The data were analyzed in the energy range from 100\,MeV to 500\,GeV, in a region of interest (ROI) of $10^o$ centering the nominal source positions.  The analysis was carried out using the maximum likelihood method (\emph{gtlike}) and standard  \emph{Fermi} {\small SCIENCE TOOLS} (version v9r12) with the instrument response function {\small `$\rm P8R2\_SOURCE\_V6$'}, Galactic diffuse model {\small `$\rm gll\_iem\_v06.fit$'} and isotropic background model {\small `$\rm iso\_p8R2\_SOURCE\_V6\_v06.txt$'}. Events which were contaminated by the bright Earths limb were excluded  using zenith angle cut of $90^o$.  Further the time bins with $TS < 9$ were excluded, which correspond to a detection significance of $\sqrt{TS}\approx 3\sigma$
We estimated monthly photon flux, energy flux, and spectral index for all the sources using the maximum likelihood analysis. 

\section{Flux distribution} 

The monthly average  $\gamma$-ray flux obtained in the analysis of $\sim$ 100 months of data were distributed  to a histogram of fluxes, for each source.   An adaptive binning was used for each source to ensure the bin width is larger than the average error of the flux within a bin.  Apart from the flux that corresponds to a TS value $TS\leq9$, we have also excluded the flux with larger  uncertainty, such that F/$\delta$F$<$2. In order to avoid the bias due to a possible lack of lower luminosity flux states, we restrict our focus only on the blazars, for which  the total excluded flux points (after the cuts mentioned above) are less than 10$\%$. After this cut, 38 (out of 50) blazars survived, which include 19 BL\,Lacs and 19 FSRQs. 

We fit all 38 flux histograms in log-scale,  with  functions 
\begin{equation}
\rm{L(x)}=\frac{1}{\sqrt{2\pi}\sigma}\exp^{\frac{-(x-\mu)^2}{2\sigma^2}} \,\,\,\, [\rm{log-normal\,distribution}]
\end{equation}
and
\begin{equation}
\rm{G(X)}=\frac{1}{\sqrt{2\pi}\sigma}\exp^{\frac{(10^x-\mu)^2}{2\sigma^2}}10^x\,log_e(10)\,\,\,\, [\rm{normal\,distribution}]
\end{equation}
where $\sigma$ and $\mu$ are the standard deviation and mean of the distribution, respectively. 

\begin{figure}
	\vspace{-2cm}
	\hspace{2.2cm}
	\begin{tabular}{c}
		\includegraphics[width=0.85\textwidth]{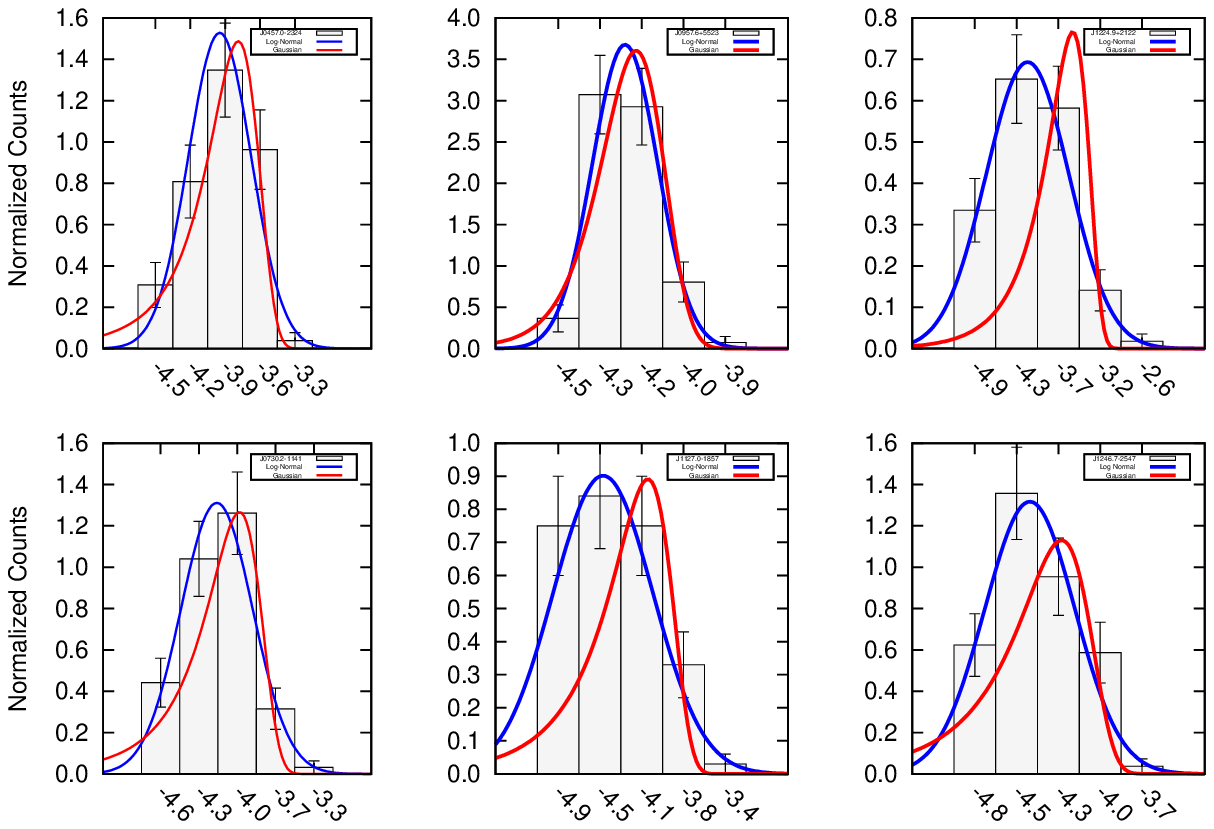} \\
		\includegraphics[width=0.85\textwidth]{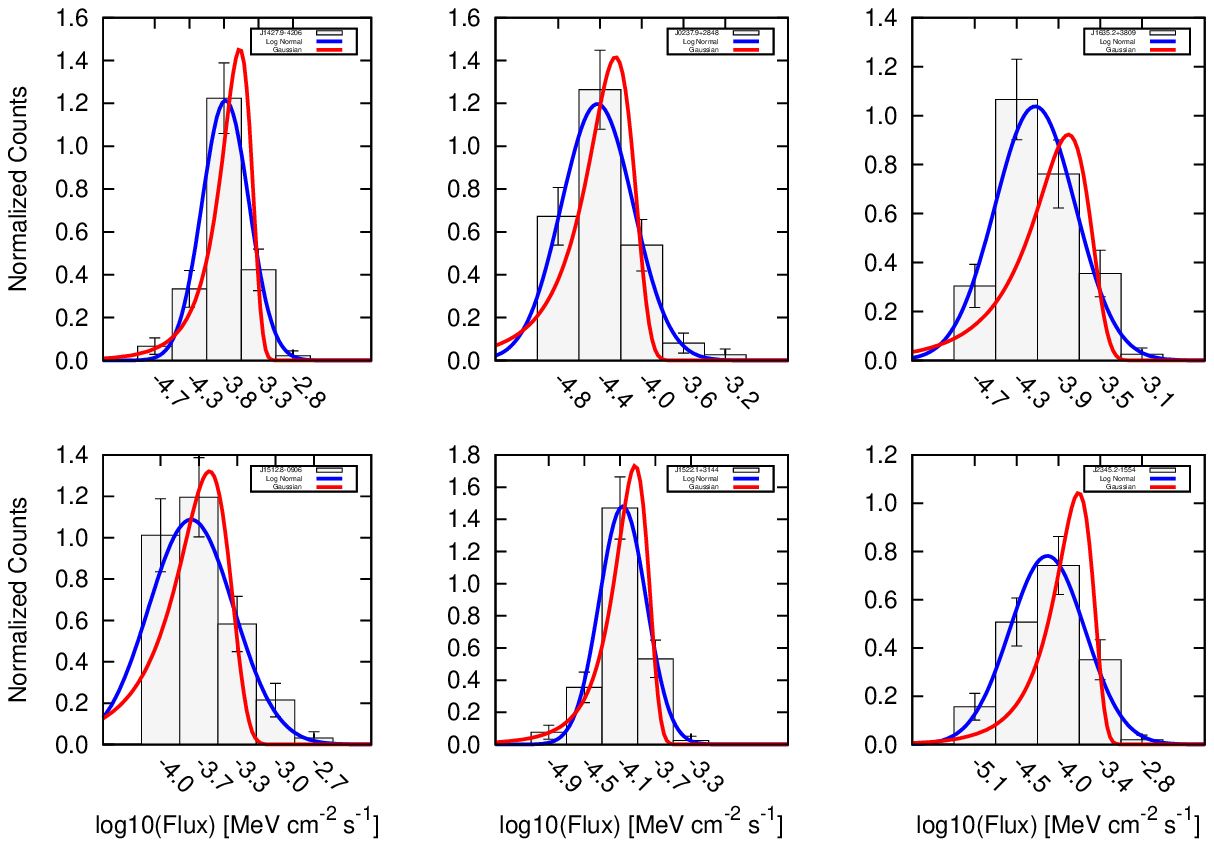} 
	\end{tabular}
    \vspace{1.5cm}
	\caption{Flux distribution of bright blazars in $\gamma$-ray band. The blue and red lines correspond to log-normal and normal fit respectively.}
	\label{fig:blazar}
\end{figure}
\setcounter{figure}{0}

\begin{figure}
	\vspace{-2cm}
	\hspace{2.2cm}
	\begin{tabular}{c}
		\includegraphics[width=0.85\textwidth]{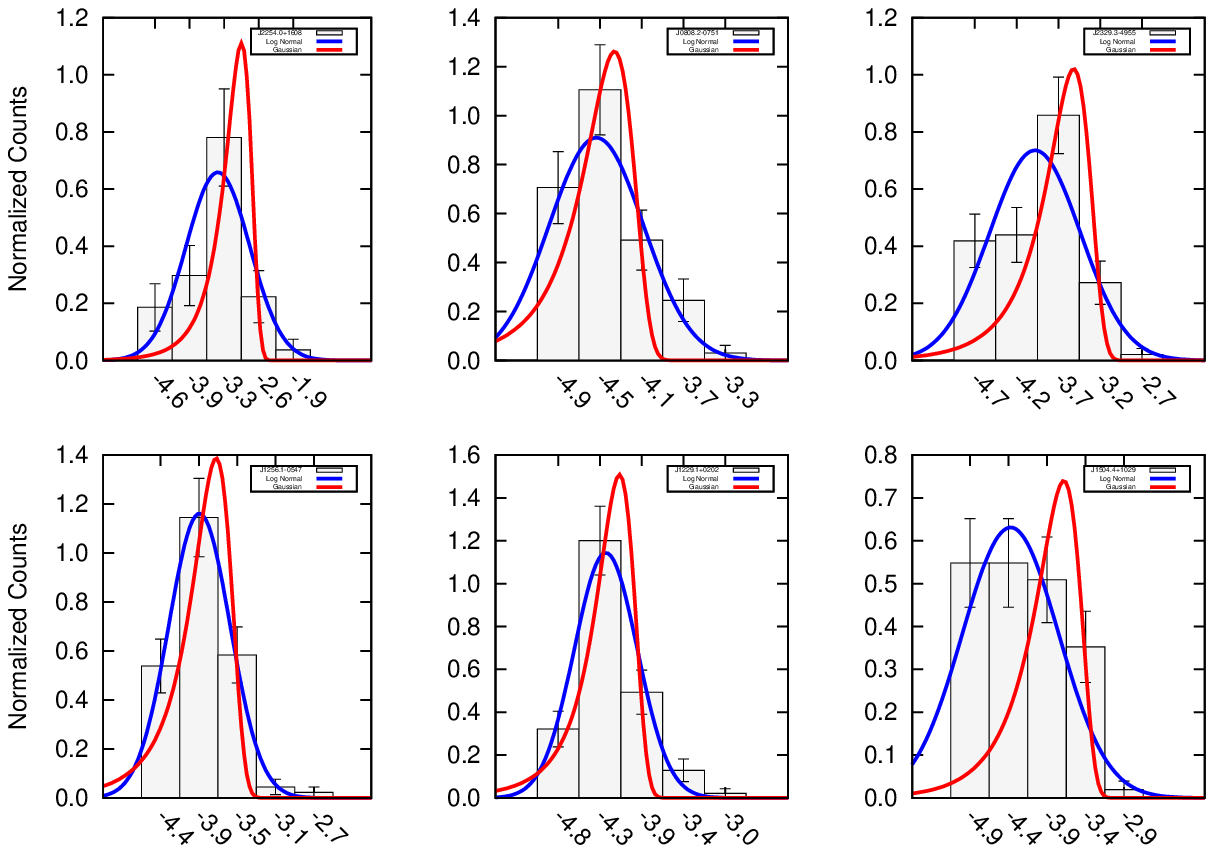} \\
		\includegraphics[width=0.85\textwidth]{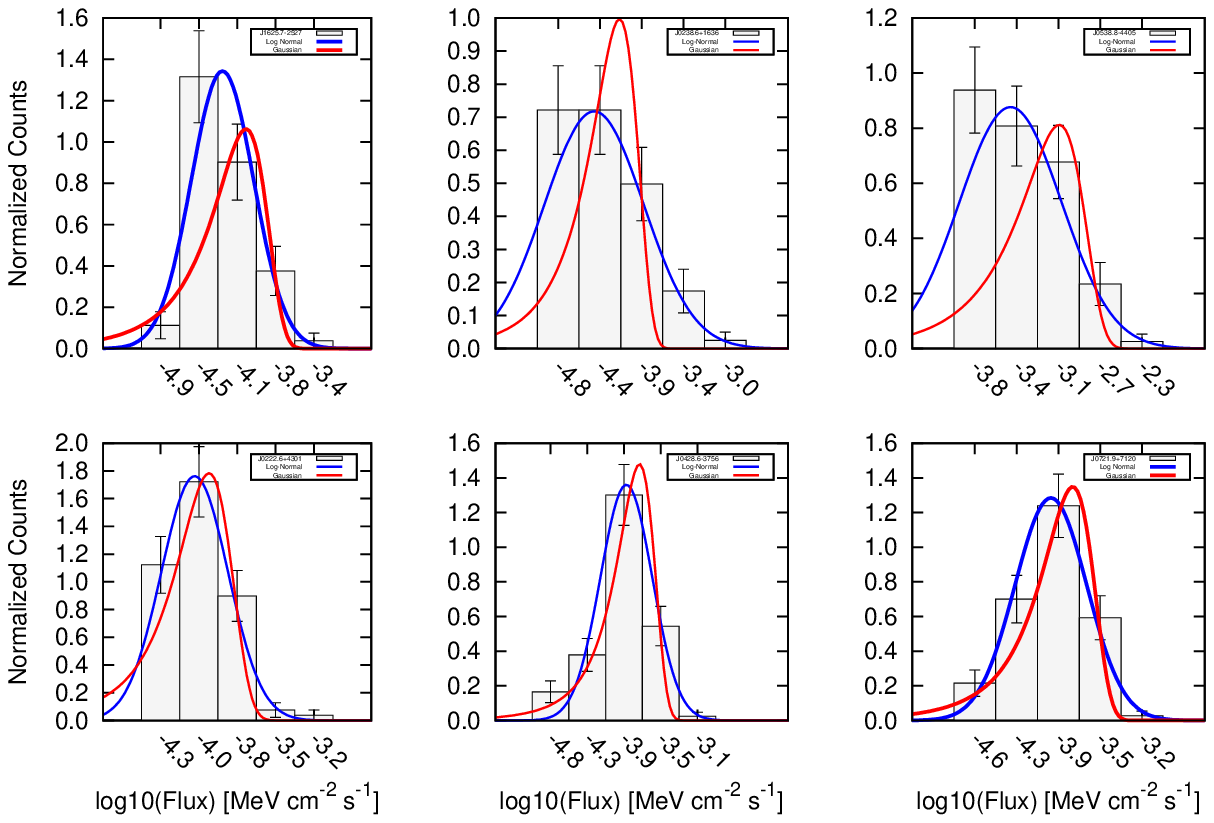} 
	\end{tabular}
	\vspace{1.5cm}
	\caption{(Continued)}
	\label{fig:blazar}
\end{figure}
\setcounter{figure}{0}

\begin{figure}
	\vspace{-2cm}
	\hspace{2.2cm}
	\begin{tabular}{c}
		\includegraphics[width=0.85\textwidth]{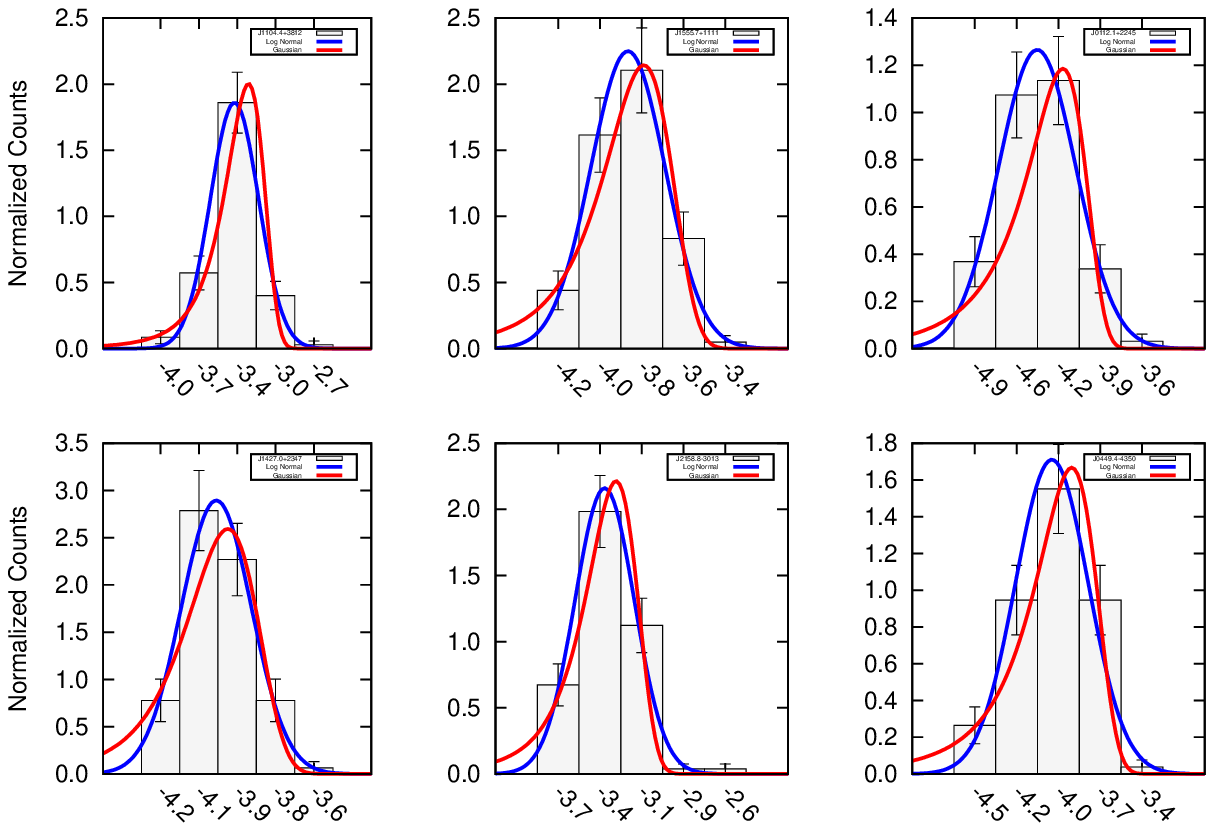} \\
		\includegraphics[width=0.85\textwidth]{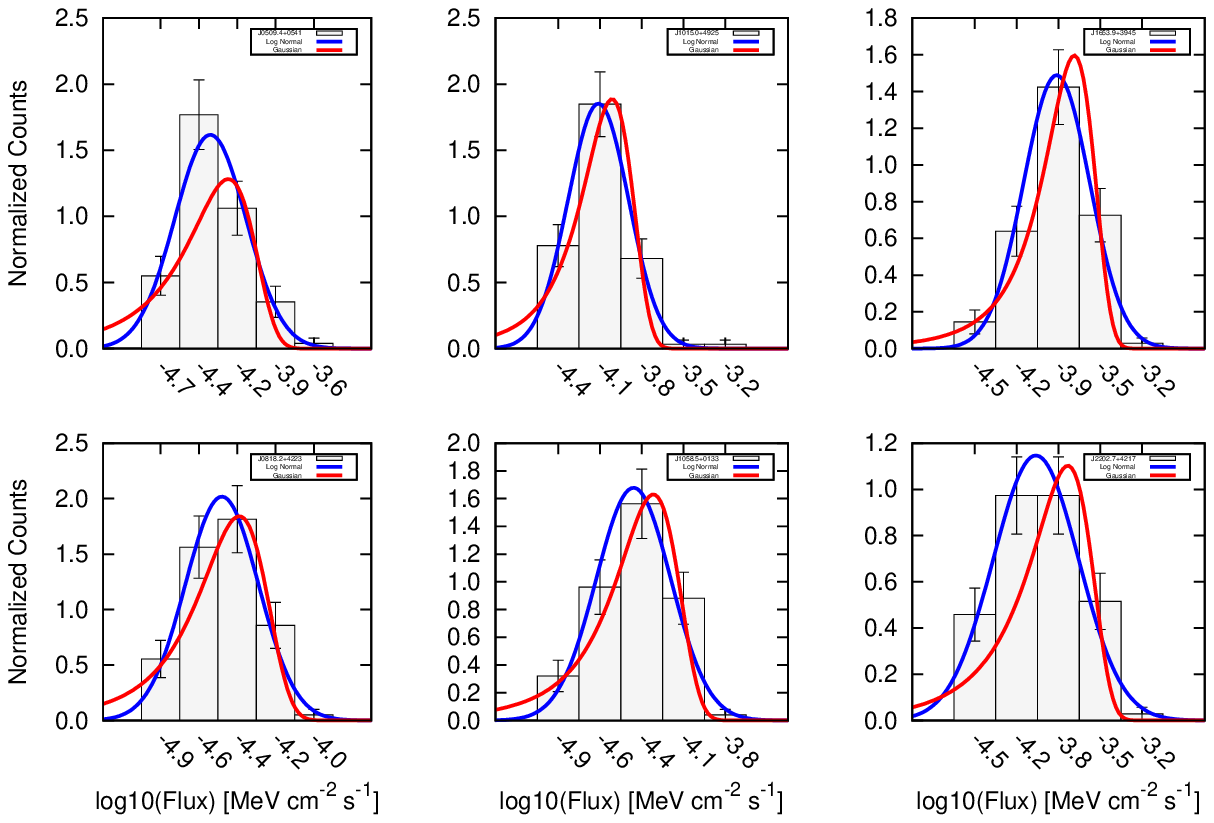} 
	\end{tabular}
	\vspace{1.5cm}
	\caption{(Continued)}
	\label{fig:blazar}
\end{figure}
\setcounter{figure}{0}

\begin{figure}
	\vspace{-2cm}
	\hspace{2.2cm}
	\begin{tabular}{c}
		\includegraphics[width=0.85\textwidth]{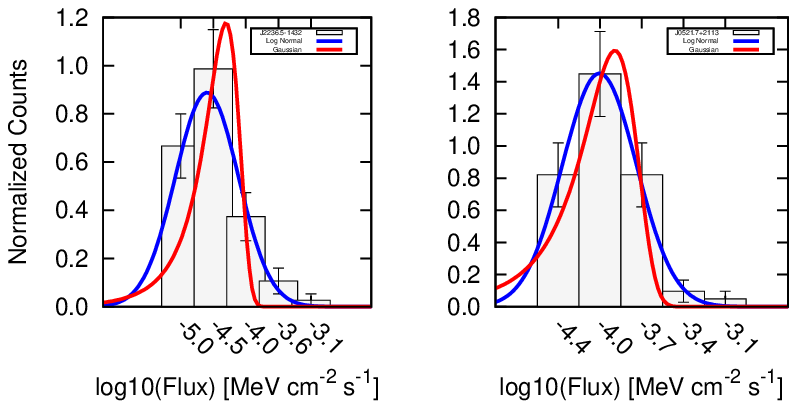} \\
	\end{tabular}
	\vspace{-1.5cm}
	\caption{(Continued)}
	\label{fig:blazar}
\end{figure}

\begin{figure}
	\centering
	\subfigure{\includegraphics[scale=0.27,angle=270]{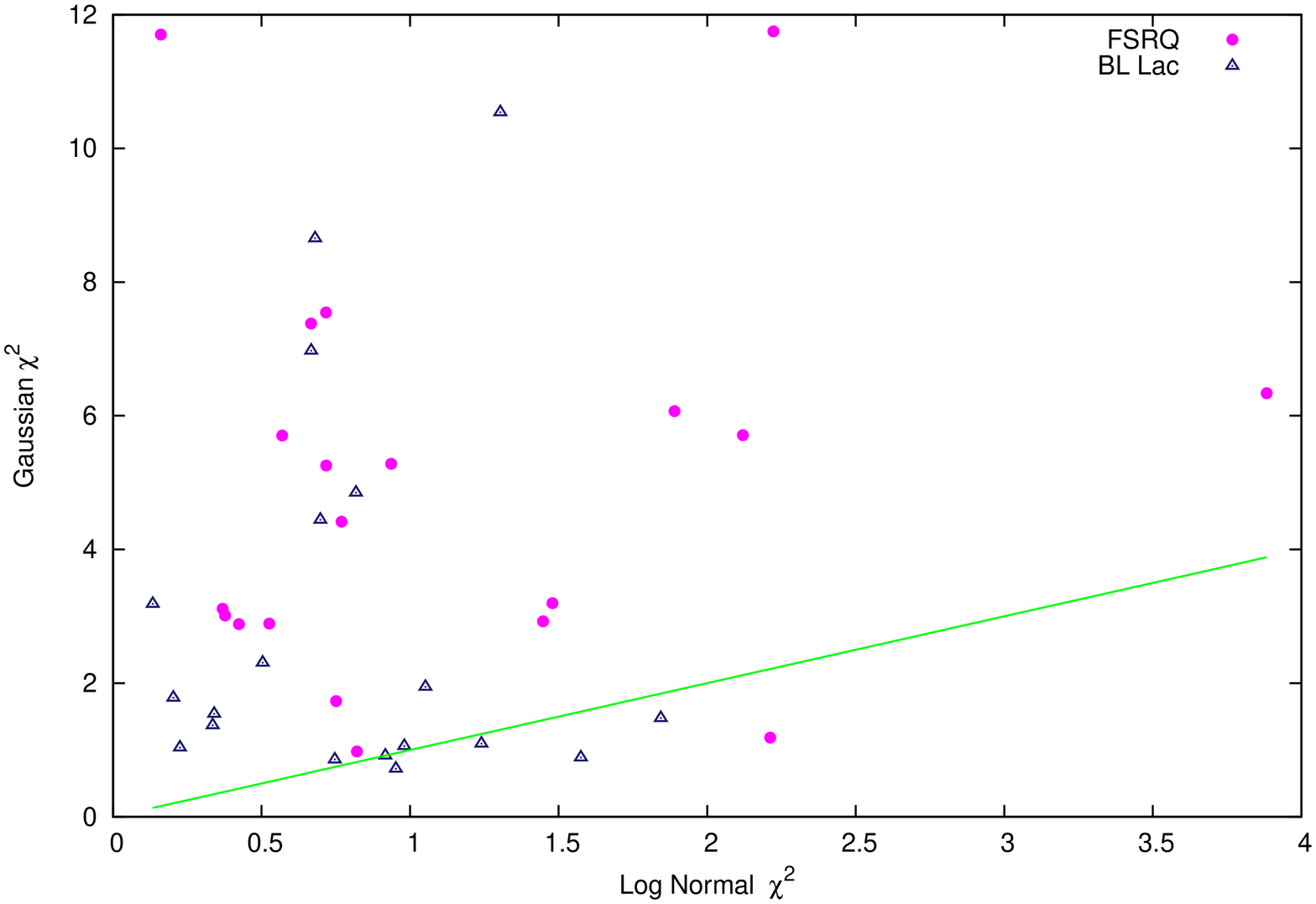}}\quad \hspace{-0.5cm}
	\subfigure{\includegraphics[scale=0.27,angle=270]{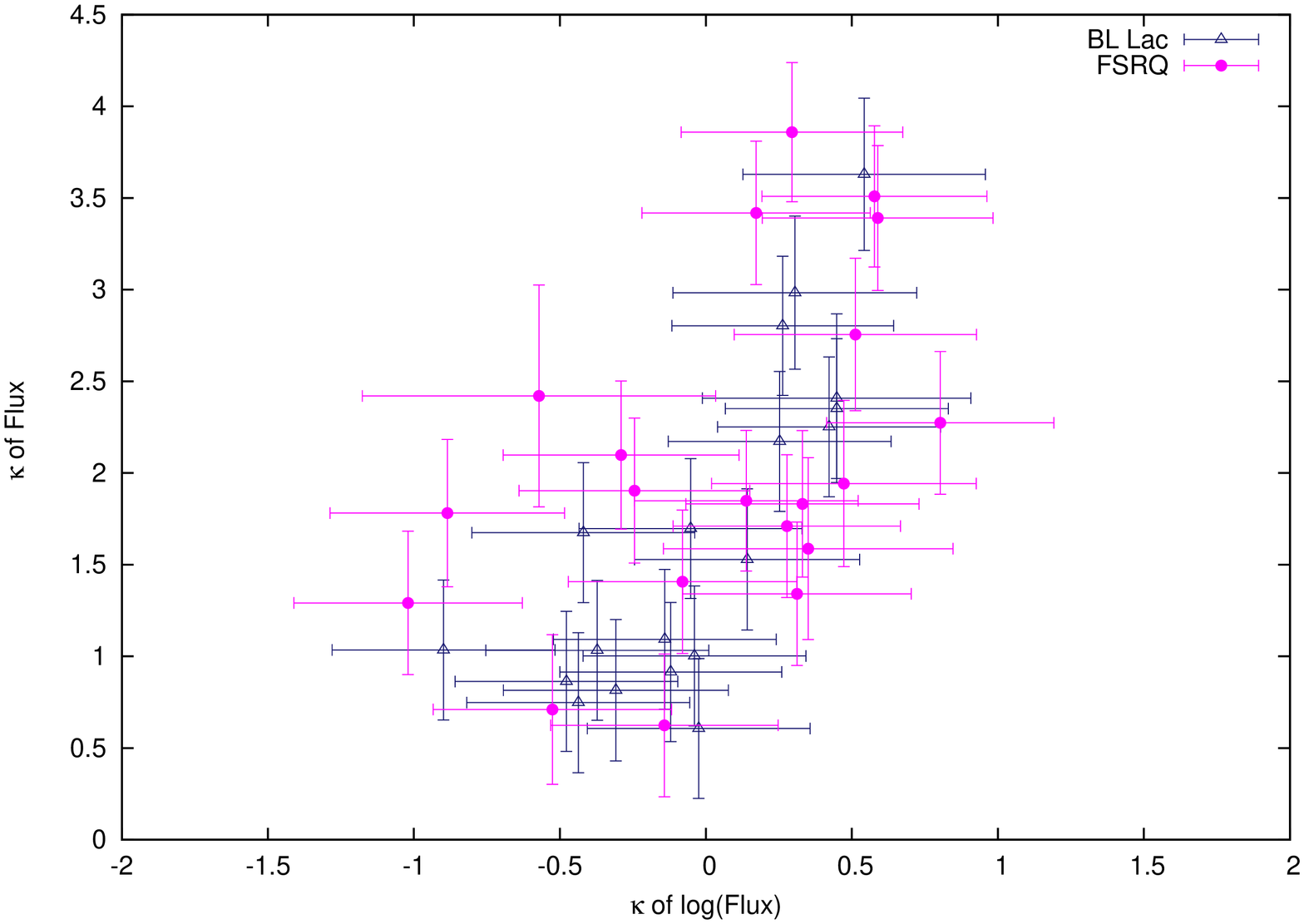}} 
	\caption{Left: the reduced $\chi^2$ obtained from the blazar flux distribution by fitting  log-normal distribution (x axis) and normal distribution (y axis). The pink circles correspond to FSRQs, while blue triangles stand for BL Lacs. The green line corresponds to the condition y=x. Right: computed skewnesss from the log-normal distribution (x axis), and normal distribution (y axis).} 
	\label{fig:skw_skw}
\end{figure}

The flux histograms of the  blazars are plotted in log-scale in Fig.\ref{fig:blazar}. The normal and log-normal fits are shown as red and blue lines respectively. Fig.\,\ref{fig:skw_skw} (left panel) shows the comparison between the reduced $\chi^2$ obtained from normal and log-normal distributions. 
The fit parameters together with the computed skewness for both distributions are shown in Table\,\ref{table:blac}. The flux distributions are found to be significantly skewed, whereas the skewness of the log of flux distribution is consistent with zero, thus suggesting log-normal trend (Fig.\,\ref{fig:skw_skw}, right panel).
Apart from calculating the reduced $\chi^2$, we have also performed Anderson-Darling (AD) test, in order to verify the normality/log-normality of the fits. The reduced $\chi^2$ of the fit, together with the AD test statistics and  rejection/null hypothesis probability (p-value) are also shown in the Table\,\ref{table:blac} . 

We note that the reduced $\chi^2$  for the  normal flux distribution  of some of the blazars  fall in a reasonable range. However, the p-value estimated from the AD  test rejects the normal  distribution ($\rm p < 0.05$) for all sources, except for the  FSRQ J0957.6+5523 (with a high reduced $\chi^2$ in this case). On the other hand, the AD test p-value, and the reduced $\chi^2$ of the  flux distribution of J0957.6+5523 do not reject the log-normal distribution of flux either.  
It is interesting to note that the  $\chi^2$ and AD tests do not reject the log-normality of the flux distribution of most of the blazars.   Nevertheless, both tests reject the  log-normality of the flux distribution of FSRQs J2329.3-4955, J1504.4+1029 and J1625.7-2527. Even though the AD tests marginally (eg: J1427.9-4206), and completely  (J1512.8-0906) reject the log-normal fits of a few blazar flux distributions, the corresponding $\chi^2$ values are reasonable enough not to reject the log-normality.

The standard deviation obtained from the log-normal fits, which is a measure of flux variability, is comparatively high for FSRQs.  In the case of BL Lacs, HBLs show a lower variability,  while the variability of LBLs are similar to that of FSRQs.  The variability of IBLs roughly fall in between HBLs and LBLs. The mean values of the standard deviation obtained from the  log-normal fit of the considered FSRQs, HBLs, IBLs and LBLs are 0.41$\pm$0.11, 0.21$\pm$0.04, 0.27$\pm$0.03, and 0.37$\pm$0.10 respectively.   We have also noticed that the standard deviation of the flux distribution of J0957.6+5523 is significantly smaller compared to other FSRQs, and could be treated as a \emph{steady FSRQ}.  

\begin{table}
	\centering
	\scriptsize
	\caption{The fitting parameters of log-normal (column: 2, 3) and normal (column: 7, 8) flux distribution of Blazars. The computed skewness (column 4 and 9), reduced $\chi^2$ (column 5 and 10), and AD statistics (column 6 and 11),  for both distributions are also shown.}
	\begin{tabular}{@{} l c c c c c c c c c c c c}
		\hline \hline
		& & & & &  FSRQ & & & & & &\\
		\hline
		Blazar  & \multicolumn{5}{c}{Log-normal}   &&  \multicolumn{5}{c}{Normal} \\   \cline{2-6} \cline{8-12}  
		Name  &  width & centroid & skewness($\rm \kappa$) & $\chi^2/dof$ &  AD (prob) &&    width$^*$ & centroid$^*$ & skewness$^*$ & $\chi^2/dof$ & AD (prob) \\
		\hline
		J0457.0-2324 &  0.26$\pm$0.03 & -3.96$\pm$0.04 & -0.52$\pm$0.40 & 2.21 & 0.64 (0.09) &&   8.33$\pm$1.40 & 11.1$\pm$1.60 & 0.71$\pm$0.40 & 1.12 & 1.50 (6.7e-04)\\
		J0730.2-1141 &  0.30$\pm$0.02 & -4.17$\pm$0.02 & -0.08$\pm$0.39 & 0.42 & 0.37 (0.41) &&   6.39$\pm$1.17 & 6.71$\pm$2.01 & 1.40$\pm$0.39 & 2.88 & 3.32 (2.3e-08)\\
		J0957.6+5523 &  0.12$\pm$0.01 & -4.24$\pm$0.01 & -0.14$\pm$0.39 & 0.53 & 0.22 (0.84) &&   1.54$\pm$0.21 & 5.85$\pm$0.28 & 0.62$\pm$0.39 & 2.89 & 0.69 (0.07122) \\
		J1127.0-1857 &  0.45$\pm$0.04 & -4.49$\pm$0.06 &  0.33$\pm$0.39 & 0.67 & 0.72 (0.06) &&   6.48$\pm$2.04 & 3.17$\pm$3.18 & 1.83$\pm$0.39 & 7.38 & 6.40 (7.7e-16)\\
		J1224.9+2122 &  0.57$\pm$0.02 & -4.17$\pm$0.03 &  0.17$\pm$0.39 & 0.16 & 0.33 (0.51) &&   23.8$\pm$8.34 & 8.26$\pm$17.6 & 3.41$\pm$0.39 & 11.7 & 10.97  ($<$2.2e-16)\\
		J1246.7-2547 &  0.30$\pm$0.03 & -4.45$\pm$0.04 &  0.31$\pm$0.39 & 0.93 & 0.72 (0.06) &&   3.79$\pm$1.08 & 3.25$\pm$1.50 & 1.34$\pm$0.39 & 5.21 & 5.44  (1.6e-13)\\
		J1427.9-4206 &  0.33$\pm$0.02 & -3.75$\pm$0.03 & -0.88$\pm$0.40 & 0.75 & 1.02 (0.01) &&   15.2$\pm$3.28 & 19.6$\pm$5.39 & 1.78$\pm$0.40 & 1.73 & 4.61 (1.6e-11)\\
		J1512.8-0906 &  0.37$\pm$0.04 & -3.73$\pm$0.05 & -0.80$\pm$0.38 & 0.71 & 2.44 (3.2e-06)  &&   15.7$\pm$4.52 & 17.6$\pm$6.85 & 2.27$\pm$0.38 & 5.26 & 10.36 ($<$2.2e-16)\\
		J0237.9+2848 &  0.33$\pm$0.02 & -4.43$\pm$0.02 &  0.59$\pm$0.39 & 0.36 & 0.71 (0.06) &&   3.04$\pm$0.77 & 3.78$\pm$1.22 & 3.39$\pm$0.39 & 3.11 & 5.44 (1.6e-13) \\
		J2254.0+1608 &  0.60$\pm$0.10 & -3.37$\pm$0.12 & -0.57$\pm$0.60 & 1.44 & 0.83 (0.03) &&   78.7$\pm$47.1 & 66.4$\pm$96.9 & 2.42$\pm$0.60 & 2.92 & 4.65 (9.7e-12)\\
		J1522.1+3144 &  0.27$\pm$0.02 & -4.07$\pm$0.02 & -1.02$\pm$0.39 & 0.84 & 0.62 (0.10) &&   5.58$\pm$0.65 & 9.13$\pm$0.96 & 1.29$\pm$0.39 & 0.90 & 2.11 (2.1e-05)\\
		J1635.2+3809 &  0.38$\pm$0.03 & -4.14$\pm$0.03 &  0.27$\pm$0.39 & 0.72 & 0.76 (0.05) &&   11.1$\pm$3.45 & 6.49$\pm$5.65 & 1.71$\pm$0.39 & 7.54 & 6.78 ($<$2.2e-16)\\
		J2329.3-4955 &  0.54$\pm$0.10 & -3.97$\pm$0.14 & -0.24$\pm$0.39 & 3.88 & 1.09 (0.01) &&   21.7$\pm$6.67 & 15.6$\pm$12.9 & 1.90$\pm$0.39 & 6.33 & 4.16 (2.0e-10) \\
		J2345.2-1554 &  0.51$\pm$0.03 & -4.10$\pm$0.04 & -0.29$\pm$0.40 & 0.56 & 0.22 (0.84) &&   14.5$\pm$5.63 & 10.7$\pm$9.85 & 2.09$\pm$0.40 & 5.70 & 6.66 ($<$2.2e-16) \\
		J0808.2-0751 &  0.45$\pm$0.05 & -4.53$\pm$0.06 &  0.51$\pm$0.41 & 0.77 & 0.65 (0.09) &&   2.72$\pm$0.78 & 2.71$\pm$1.25 & 2.75$\pm$0.41 & 4.42 & 8.46 ($<$2.2e-16)\\
		J1229.1+0202 &  0.34$\pm$0.04 & -4.28$\pm$0.04 &  0.57$\pm$0.38 & 1.48 & 0.70 (0.06) &&   3.89$\pm$1.25 & 5.29$\pm$1.93 & 3.50$\pm$0.38 & 3.12 & 12.89 ($<$2.2e-16)\\
		J1256.1-0547 &  0.34$\pm$0.02 & -3.94$\pm$0.02 &  0.29$\pm$0.37 & 0.38 & 0.27 (0.68) &&   9.92$\pm$2.83 & 11.9$\pm$4.16 & 3.86$\pm$0.37 & 3.01 & 8.63 ($<$2.2e-16)\\
		J1504.4+1029 &  0.63$\pm$0.10 & -4.38$\pm$0.15 &  0.14$\pm$0.38 & 2.22 & 0.98 (0.01) &&   17.9$\pm$7.48 & 5.29$\pm$11.4 & 1.84$\pm$0.38 & 11.7 & 8.54 ($<$2.2e-16)\\
		J1625.7-2527 &  0.30$\pm$0.04 & -4.28$\pm$0.05 &  0.47$\pm$0.45 & 2.12 & 1.11 (0.01) &&   5.96$\pm$1.88 & 4.64$\pm$2.98 & 1.94$\pm$0.45 & 5.71 & 6.87 ($<$2.2e-16)\\
		\hline
		& & & & &  BL Lac & & & & & &\\
		\hline
		J0222.6+4301  & 0.23$\pm$0.02 & -4.04$\pm$0.02 &  0.45$\pm$0.38 & 0.50 & 0.44 (0.28)   &&   5.23$\pm$1.07 & 8.88$\pm$1.42 & 2.35$\pm$0.38 & 2.30 & 3.55 (6.1e-09)  \\
		J0238.6+1636  & 0.56$\pm$0.06 & -4.42$\pm$0.08 &  0.30$\pm$0.42 & 0.67 & 0.85 (0.03)   &&   5.19$\pm$2.04 & 3.57$\pm$2.97 & 2.98$\pm$0.42 & 6.97 & 7.70  ($<$2.2e-16)  \\
		J0428.6-3756  & 0.33$\pm$0.03 & -3.95$\pm$0.04 & -0.90$\pm$0.38 & 1.84 & 11.45 (0.01)   &&   9.75$\pm$2.14 & 11.5$\pm$2.83 & 1.03$\pm$0.38 & 1.47 & 2.15  (1.7e-05)  \\
		J0538.8-4405  & 0.45$\pm$0.06 & -3.48$\pm$0.08 &  0.42$\pm$0.38 & 1.30 & 1.02 (0.01)   &&   73.1$\pm$27.3 & 30.5$\pm$40.2 & 2.25$\pm$0.38 & 10.5 & 8.04  ($<$2.2e-16)  \\
		J0721.9+7120  & 0.31$\pm$0.02 & -3.96$\pm$0.03 & -0.37$\pm$0.38 & 1.05 & 0.66 (0.08)   &&   9.66$\pm$1.75 & 11.1$\pm$2.56 & 1.03$\pm$0.38 & 1.94 & 2.25  (9.7e-06)  \\
		J1104.4+3812  & 0.21$\pm$0.02 & -3.39$\pm$0.02 & -0.42$\pm$0.38 & 0.91 & 0.48 (0.22)   &&   23.0$\pm$1.89 & 45.1$\pm$3.15 & 1.67$\pm$0.38 & 0.92 & 2.67  (8.9e-07) \\
		J1427.0+2347  & 0.14$\pm$0.01 & -4.00$\pm$0.01 & -0.02$\pm$0.38 & 0.20 & 0.34 (0.50)   &&   3.74$\pm$0.56 & 9.92$\pm$0.68 & 0.61$\pm$0.38 & 1.78 & 1.13  (0.01)  \\
		J1555.7+1111  & 0.17$\pm$0.01 & -3.84$\pm$0.01 & -0.04$\pm$0.38 & 0.22 & 0.5124 (0.19)   &&   6.88$\pm$0.78 & 14.6$\pm$0.97 & 1.00$\pm$0.38 & 1.04 & 2.12 (2.1e-05)  \\
		J2158.8-3013  & 0.18$\pm$0.01 & -3.38$\pm$0.02 &  0.25$\pm$0.38 & 0.75 & 0.1918 (0.89)   &&   19.1$\pm$2.52 & 42.2$\pm$2.75 & 2.17$\pm$0.38 & 0.86 & 2.68 (8.4e-07)  \\
		J0112.1+2245  & 0.31$\pm$0.01 & -4.42$\pm$0.01 & -0.05$\pm$0.38 & 0.13 & 0.14 (0.97)   &&   4.06$\pm$0.76 & 3.50$\pm$1.31 & 1.70$\pm$0.38 & 3.19 & 4.21  (9.2e-11)  \\
		J0449.4-4350  & 0.23$\pm$0.02 & -4.02$\pm$0.03 & -0.44$\pm$0.38 & 1.57 & 0.51 (0.20)   &&   6.35$\pm$0.80 & 9.55$\pm$0.93 & 0.75$\pm$0.38 & 0.89 & 1.11 (0.01)  \\
		J0509.4+0541  & 0.25$\pm$0.02 & -4.36$\pm$0.02 &  0.14$\pm$0.39 & 0.69 & 0.19 (0.89)   &&   3.63$\pm$0.95 & 3.65$\pm$1.27 & 1.53$\pm$0.39 & 4.44 & 3.02 (1.2e-07) \\
		J0818.2+4223  & 0.20$\pm$0.01 & -4.52$\pm$0.02 & -0.12$\pm$0.38 & 0.95 & 0.26 (0.72)   &&   1.61$\pm$0.16 & 3.02$\pm$0.18 & 0.91$\pm$0.38 & 0.72 & 2.05 (3.0e-05)  \\
		J1015.0+4925  & 0.22$\pm$0.01 & -4.09$\pm$0.01 &  0.26$\pm$0.38 & 0.34 & 0.31 (0.56)   &&   4.45$\pm$0.66 & 8.11$\pm$0.97 & 2.80$\pm$0.38 & 1.36 & 3.94 (7.1e-10)  \\
		J1058.5+0133  & 0.25$\pm$0.03 & -4.43$\pm$0.04 & -0.31$\pm$0.39 & 1.24 & 0.45 (0.27)   &&   2.70$\pm$0.43 & 3.83$\pm$0.41 & 0.82$\pm$0.39 & 1.09 & 2.67 (8.6e-07)  \\
		J1653.9+3945  & 0.28$\pm$0.02 & -2.88$\pm$0.03 & -0.48$\pm$0.38 & 0.98 & 0.63 (0.10)   &&   9.33$\pm$1.22 & 13.6$\pm$1.58 & 0.86$\pm$0.38 & 1.06 & 1.30 (2.1e-3) \\
		J2202.7+4217  & 0.35$\pm$0.03 & -4.02$\pm$0.04 & -0.14$\pm$0.38 & 0.81 & 0.74 (0.05)   &&   11.6$\pm$3.02 & 9.69$\pm$4.56 & 1.09$\pm$0.38 & 4.85 & 2.82 (3.8e-07)  \\
		J2236.5-1432  & 0.41$\pm$0.03 & -4.51$\pm$0.04 &  0.54$\pm$0.42 & 0.68 & 0.63 (0.10)   &&   7.91$\pm$2.48 & 2.23$\pm$4.92 & 3.63$\pm$0.42 & 8.65 & 12.24 ($<$2.2e-16) \\
		J0521.7+2113  & 0.28$\pm$0.02 & -4.05$\pm$0.02 &  0.45$\pm$0.46 & 0.34 & 0.38 (0.39)   &&   6.06$\pm$1.26 & 8.30$\pm$1.66 & 2.41$\pm$0.46 & 1.54 & 3.69 (2.6e-09)  \\
		\hline \hline
		\multicolumn{10}{l}{$^*$units of $ 10^{-5}$\,MeV\,cm$^{-2}$\,s$^{-1}$} \\
	\end{tabular}
	\label{table:blac}
\end{table} 

\subsection{The case of Uncertain type blazars}

More than 500 sources were classified as blazars of uncertain type (BCU) in 3LAC.  Even though these  sources are  associated  with  extra-galactic counterparts, and  show some of the blazar characteristics,  they lack reliable classification based on spectral information. In order to  investigate the flux distribution properties of such sources, we analyzed the long term data (in the same time period as of the other bright blazars) of three brightest BCUs (namely, J0522.9-3628, J0532.0-4827, and J1328.9-5608). These sources are comparatively less brighter than the known classified blazars that we considered. After analyzing the \emph{Fermi}-LAT data of three bright BCUs with the standard \emph{Fermi} {\small SCIENCE TOOL} and using the quality cuts i.e. the flux after the cuts of $F/\delta F<2$ and $TS\leq9$,  it was not possible to match with the acceptance criteria of 90\%. Therefore, we modified our acceptance criteria from 90$\%$ to 60$\%$ of total flux. Since the flux states with TS$<$9 belong to low flux states (quiescent states) or less variable states,  the difference in the acceptance criteria will not significantly bias our final results. The flux distribution of all three sources suggest  log-normal distributions. The reduced $\chi^2$ of the log-normal flux distribution for the sources J0522.9-3628, J0532.0-4827, and J1328.9-5608 are found to be 0.29, 1.18 and 0.73 respectively (instead of 5.67, 3.09, 5.79 in the case of normal distribution). The AD test statistic p-values of the logarithmic  flux distribution for the sources (in the same order as above) are  0.78, 0.80 and 0.17 (instead of  $6.72\times10^{-15}$, $2.86\times10^{-16}$ and $2.37\times10^{-11}$ for normal distribution), which also propose log-normal distributions of flux over normal distributions. The obtained standard deviation from the log-normal fit of J0522.9-3628 and J1328.9-5608 are 0.29 and 0.34 respectively, while it is high (0.47, which is similar to FSRQs and LBLs) for the source  J0532.0-4827. 

\begin{figure}
	\vspace{-3.0cm}
	\hspace{2.2cm}
	\includegraphics[width=1.0\textwidth]{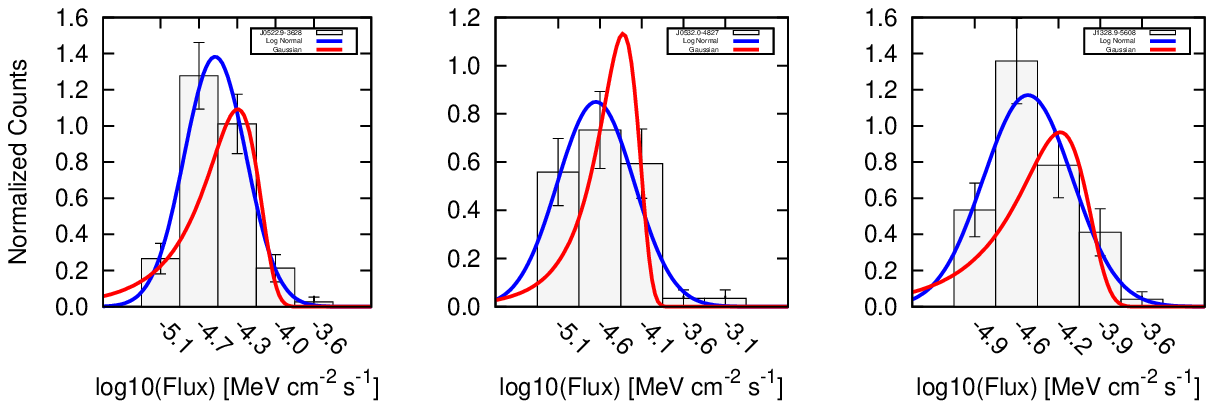} \\  
	\vspace{-1.5cm}
	\caption{Flux distribution of bright blazars in $\gamma$-ray band. The pointers are same as in the Fig.\,\ref{fig:blazar}}
	\label{fig:ucb}
\end{figure}

\section{Association of flux distribution with spectral index}

In order to investigate the association of flux variability  with spectral indices, we estimated the monthly spectra of the blazars. All the considered blazars were well described by either simple power law or log parabola models. In order to compare the spectral indices of the spectra described by both models, it would be meaningful to calculate the spectral indices at a specific energy, which was chosen to be 1 GeV  (denoted by $\alpha_{1\rm{GeV}}$). The spectral index at energy E is defined by
\begin{equation}
\alpha_E=\alpha+2\beta\log(E/E_p)
\end{equation}
where $\alpha$ is spectral index at pivot Energy $\rm E_p$, and $\beta$ is the measure of spectral curvature.
We have found that the average spectral indices for all FSRQs is 2.28$\pm$0.03.  However, the FSRQ J2254.0+1608 (a.k.a. 3C 454.3, which is the brightest blazar in $\gamma-ray$ band) shows a harder spectrum of index 1.72$\pm$0.06. It has to be also noted that the spectral index of the \emph{steady FSRQ} J0957.6+5523  is also smaller, which is 1.91$\pm$0.05. The mean spectral indices of all other FSRQs fall in the  range of 2.08-2.75. However, the mean spectral indices of BL\,Lacs fall in the range of 1.61-2.49, with a mean value of 2.01$\pm$0.11.

The mean spectral indices of BCUs J0522.9-3628,  J0532.0-4827 and  J1328.9-5608 are 2.70, 2.64, and 2.74 respectively. These values fall in the range of spectral indices of other bright FSRQs, LBLs or IBLs, but comparatively higher than the spectral indices of HBLs. 

\begin{figure}
	\centering
	\includegraphics[scale=0.40, angle=270]{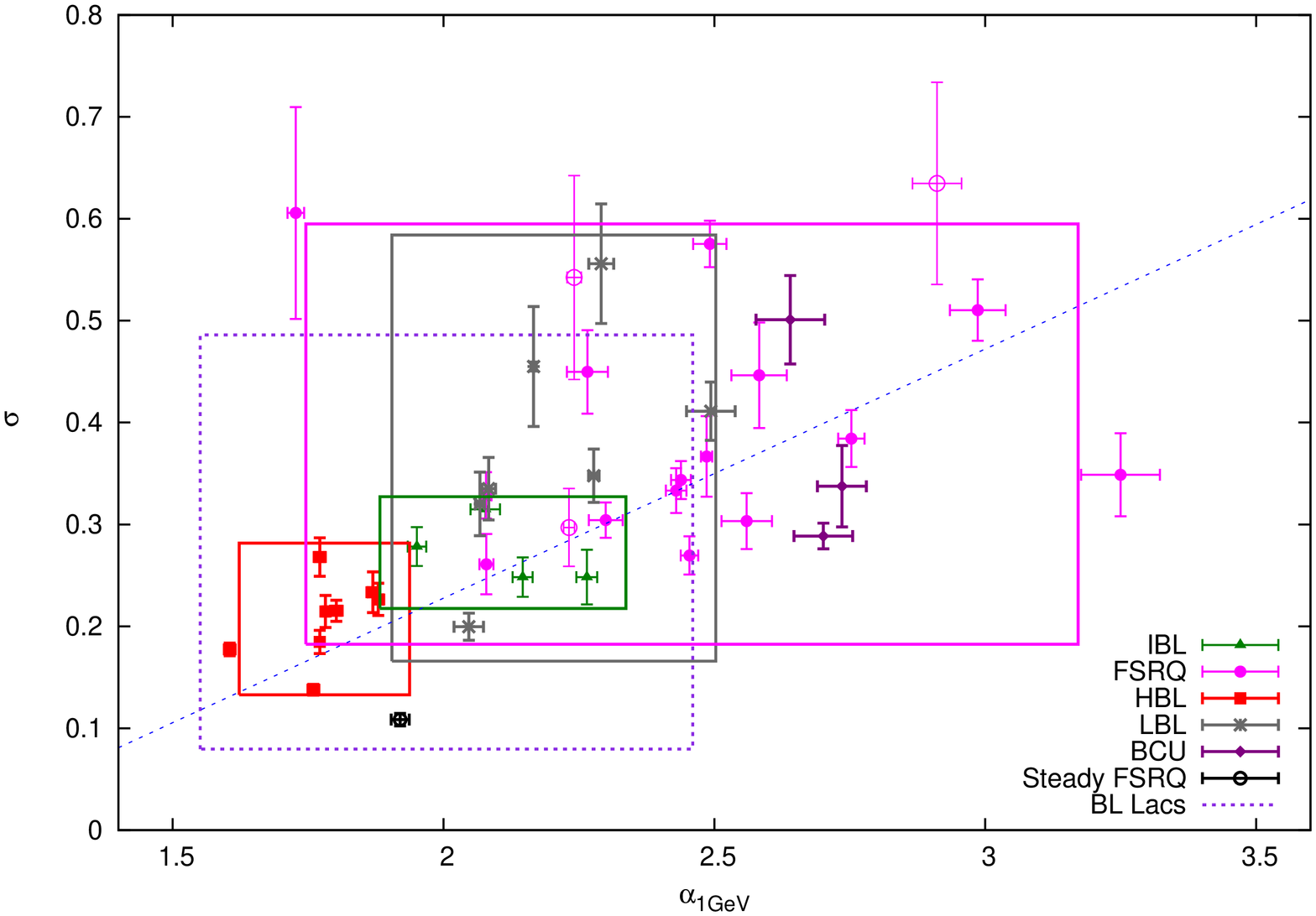}
	\vspace{-0.5cm}
	\caption{The standard deviation ($\sigma$) of the blazar flux distribution vs their corresponding average spectral index at 1\,GeV ($\alpha_{1\,\rm{GeV}}$). Each blazar  class has been marked with different pointers. HBL: red filled square, IBL: green triangle, LBL: gray dumbbell, FSRQ: pink circle, BCU: purple tilted square. The FSRQs which reject the log-normal flux distribution has been marked as pink open circle. The \emph{steady FSRQ} has been shown separately black filled circle. The rectangle boxes correspond to two-standard-deviation uncertainty from the mean of $\sigma$ and $\alpha_{1\,\rm{GeV}}$ for each blazar class. The color of the box is chosen to be same as that of pointers for each blazar class. The blue-violet dotted box corresponds to two-standard-deviation uncertainty from the mean of $\sigma$ and $\alpha_{1\,\rm{GeV}}$ for all BL\,Lac sources} 
	\label{fig:lum_sig}
\end{figure}

The flux variability has been plotted against the spectral index in Fig\,\ref{fig:lum_sig}. The HBLs (red square) fall in the left-low corner of the diagram, while FSRQs (pink circle) show a wider distribution. The IBLs (green triangle) and LBLs (gray dumbbell) fall in similar range of spectral index, though the variability distribution is wider in the case of former. The boxes represent the two-standard deviation uncertainty in $\alpha_{1\,GeV}$ and $\sigma$ from their respective mean value. The bright BCUs that were analyzed fall in the band of FSRQs. We also notice a slight correlation between the spectral index and standard deviation of the flux. The $\alpha_{1\,GeV}$ and $\sigma$ of all blazars can be roughly fitted by a straight line of slope 0.24$\pm$0.04,  which is indicated by a blue dotted line  in the figure\,\ref{fig:lum_sig}. 
The brightest BCUs show similar features as in the case of other blazars with respect to spectral index and flux distribution. All three BCUs (indicated as purple tilted square in fig.\,\ref{fig:lum_sig}) fall beyond the 2-standard deviation uncertainty region (dotted box) of BL\,Lac sources. However, they are placed within the 2-standard deviation uncertainty region of FSRQ.

Additionally, we have also compared the monthly spectral indices of each source with their corresponding luminosity. It is interesting to note that every  source shows \emph{harder when brighter} phenomena.
The  monthly spectral indices and corresponding luminosity can be well fitted by a straight line. 
The average indices are found to be -0.09+0.01 for FSRQs, while it is -0.11$\pm$0.01, -0.13$\pm$0.01,  -0.13$\pm$0.01 for HBL, IBL, and LBL respectively, suggesting the \emph{harder when brighter} phenomena among the different bright blazar classes are not significantly different.

\section{Discussion}

After a detailed study on the $\gamma$-ray flux of 38 brightest blazars, we found that the flux distributions predominantly suggest  log-normal distribution rather than a normal distribution. We verified the log-normality (over normality) using both reduced $\chi^2$ and AD-test. The log-normality was rejected only in the case of three (in a sample of 38) blazars. However, the normal distribution was  rejected for all blazars (except J0957.6+5523, though the reduced $\chi^2$ was high). The flux distribution of the three brightest BCUs follow  log-normal distribution. From the obtained spectral index and the flux standard deviation parameters, they fall beyond the 2-standard deviation uncertainty limits of HBL, IBL and LBL. Though it can not be asserted, we are tempted to associate these sources with FSRQs.

The log-normal distribution of the observed flux indicates the perturbation associated with the emission process to be of
multiplicative nature rather than additive \citep{Lyubarskii1997,Arevalo2006}. Flux variation in blazars can be attributed to the complex interplay between the intrinsic and source parameters. A simple scenario is to associate the flux variation with the fluctuation in the emitting electron
number density or the magnetic field. However, the linear dependence of the these quantities with the differential flux suggests, this will cause a normal
flux distribution contrary to the observations. Alternatively, the particle acceleration and the diffusion processes can modify the shape of the emitting
electron distribution \citep{Kirk1998} and hence can be accounted for various flux distributions, including a log-normal one. The flux variation
can also be associated with the change in the emission region geometry. Even though
the change in volume associated with this can only produce normal flux distribution, inclusion of
light travel time effects can significantly modify the same \citep{Chiaberge1999}.
However, the timescales associated with these processes are too short and hence will not
reflect the log-normal distribution obtained in our study, where we used  monthly averaged fluxes.

A log-normal flux distribution can directly hint the linkage  of blazar jet
with the accretion phenomena since the latter is well proven to produce such
distribution through the study of galactic X-ray binaries (XRBs)\citep{Uttley2001}. The fluctuations in the disk at different radii are known to be produced independently  by viscosity fluctuation on  local viscous time scales, which modulates the mass accretion rate at larger distances from black-hole. The accretion rate variations then propagate to small radii through accretion flow and concoction of variations at different radii results in multiplicative emission. This model was put forward by \cite{Lyubarskii1997} for explanation of observed X-ray variability time-scales in XRBs. Also, for non beamed accreting objects the variability timescales are found to be  proportional to $\rm M/\dot{m}$, where M is mass of black hole and $\rm \dot{m}$ is accretion rate \citep{Kording2007}.  \cite{McHardy2008}  had found that same relation surprisingly holds even for beamed jet emission  from blazars e.g, 3C\,273, which should have otherwise shorter observed variability timescale due to relativistic time dilation than the timescale predicted using the black hole mass and accretion rate.  Consequently, this lead to inference  that source of variations in blazars lie out-side the jet i.e, in the accretion disks which then modulates the jet emission. A detailed study of month scale averaged flux distribution of blazars can hence be a key to understand disk-jet connection. 

On the
contrary to the interpretations above, a log-normal flux distribution can also arise from
additive processes under specific conditions. For example, if the blazar jet is
assumed to be a large collection of mini-jets, then the logarithm of composite
flux will show a normal distribution \citep{Biteau2012}.

We note that the AD statistics does not reject the normality of flux distribution of J0957.6+5523. Moreover, the standard deviation obtained from the flux distribution of this source exhibits a significant difference from that of other blazars. 3LAC labeled this source as an FSRQ, based on  the presence of broad optical emission lines, large redshift and high $\gamma$-ray luminosity of the order of $\rm \approx 10^{47} erg\,s^{-1}$. However, the integrated spectrum and morphological properties obtained from the VLBA observations question the FSRQ classification of the source, and suggest it as one of the weakest Compact Symmetric Object \citep{Rossetti2005}.  Moreover, the brightness temperature  of this source was found to be significantly lower ($2\times10^{8}$\,K at 5\,GHz, \citealt{Taylor2007}),  than that of  other  $\gamma$-ray blazars \citep{McConville2011}.  These studies, together with our results, suggest that more multi-wavelength studies are required before  associating this source to an FSRQ.

\begin{figure}
	\minipage{0.25\textwidth}
	\includegraphics[width=\linewidth, angle=270]{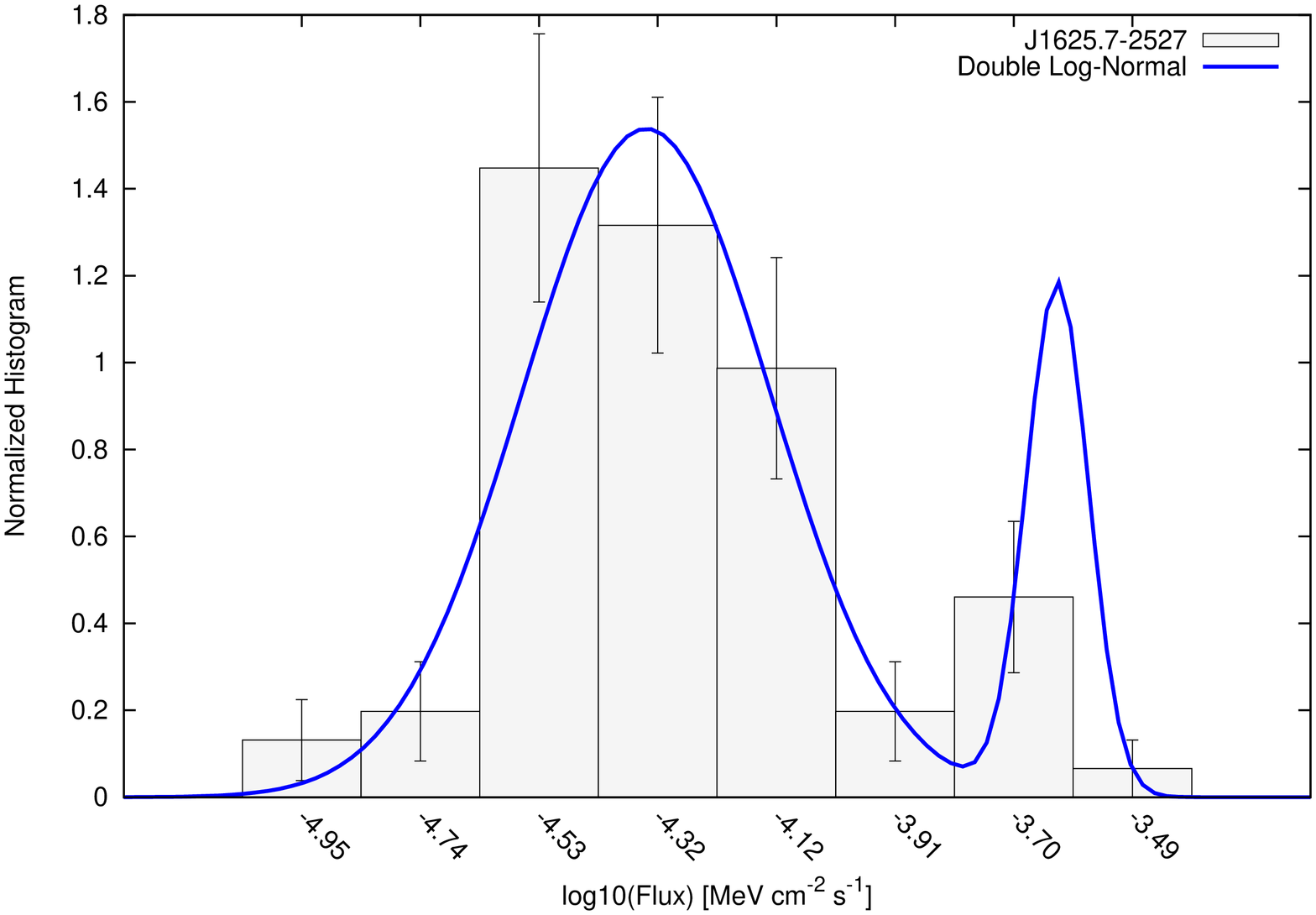}\hspace{-2.0cm}
	\endminipage\hfill
	\minipage{0.25\textwidth}
	\includegraphics[width=\linewidth, angle=270]{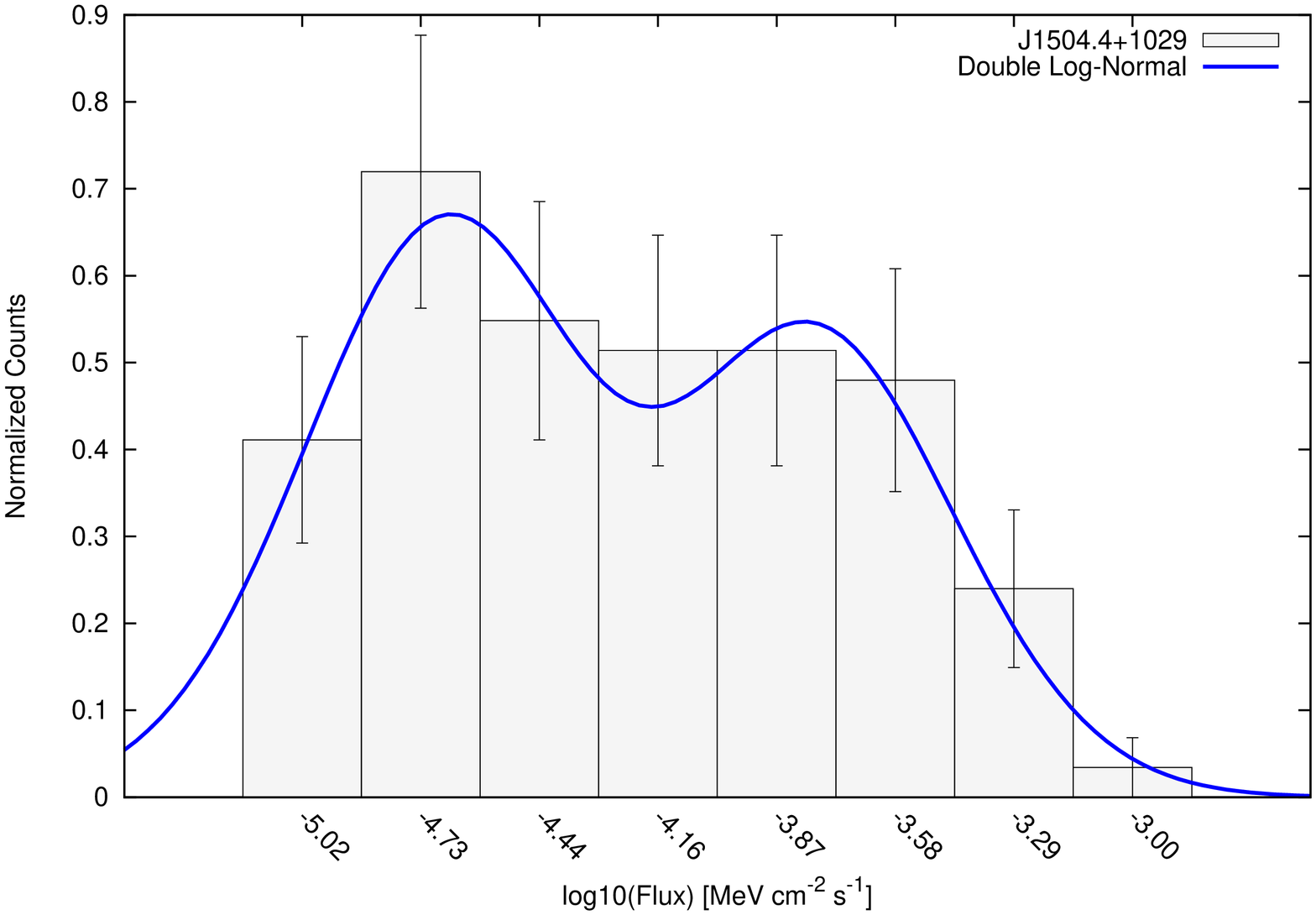}\hspace{-2.0cm}
	\endminipage\hfill
	\minipage{0.25\textwidth}%
	\includegraphics[width=\linewidth, angle=270]{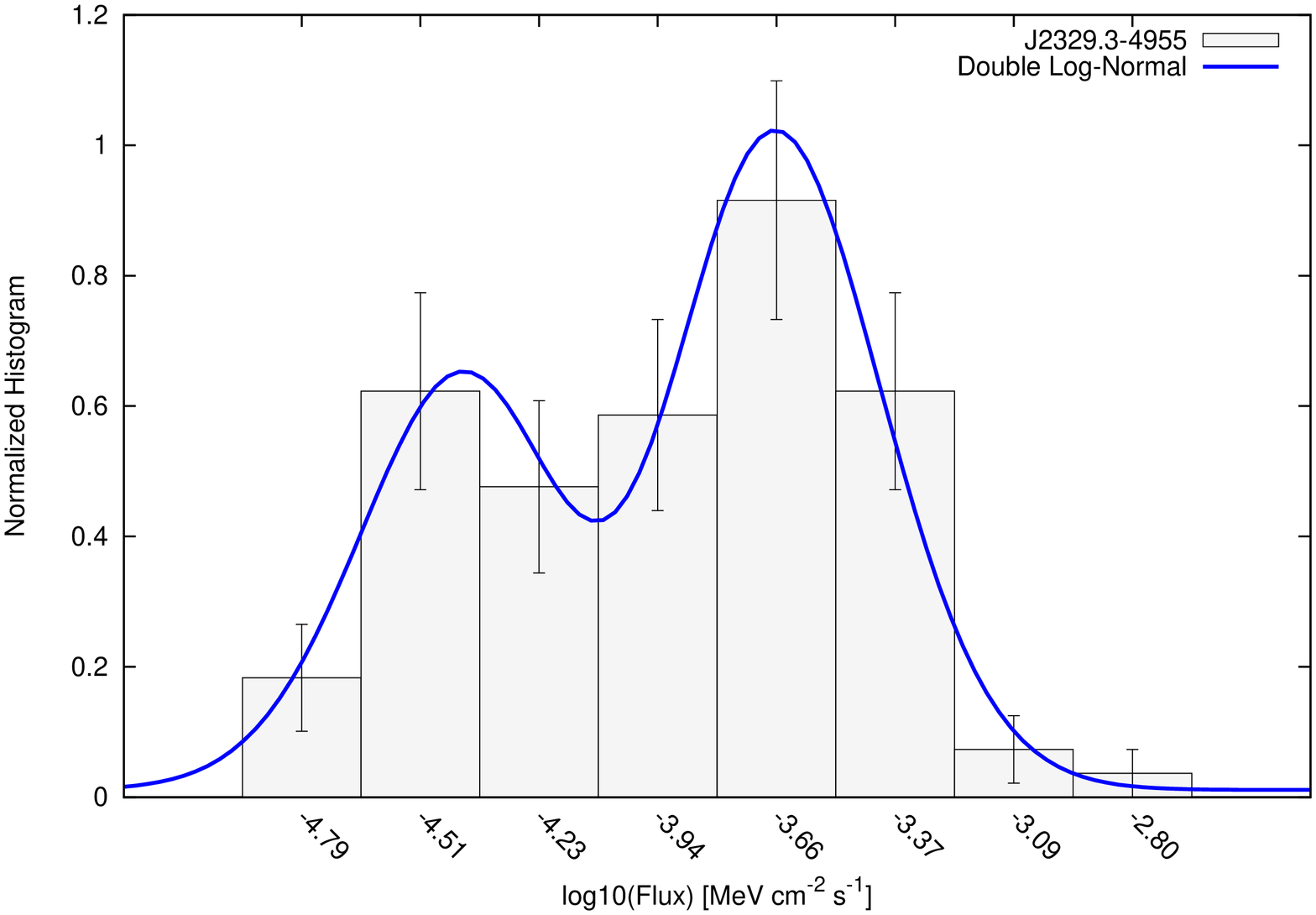}\hspace{-2.0cm}
	\endminipage
	\caption{Double log-normal fit to the flux distribution of the blazars J1625.7-2527, J1504.4+1029, and J2329.3-4955, which reject log-normal distribution. Since the statistics is not significant enough, these fits should be taken only as indicative.}\label{fig:double}
\end{figure}

We have also investigated the possibility of double log-normal distribution of fluxes for the sources that reject log-normal distribution both in $\chi^2$ and  AD test (J1625.7-2527, J1504.4+1029, and J2329.3-4955). The two distinct log-normal profiles may indicate  different flux states correspond to a low and high states of the source \citep{Kushwaha2016}. It is interesting to note that the flux distribution of all three sources exhibit hints of double log-normal distribution (Fig.\,\ref{fig:double}). However, the statistics of this distribution is not significant enough,  hence these result should be taken only as an indicative.

\begin{figure}
	\centering
	\subfigure{\includegraphics[scale=0.27,angle=270]{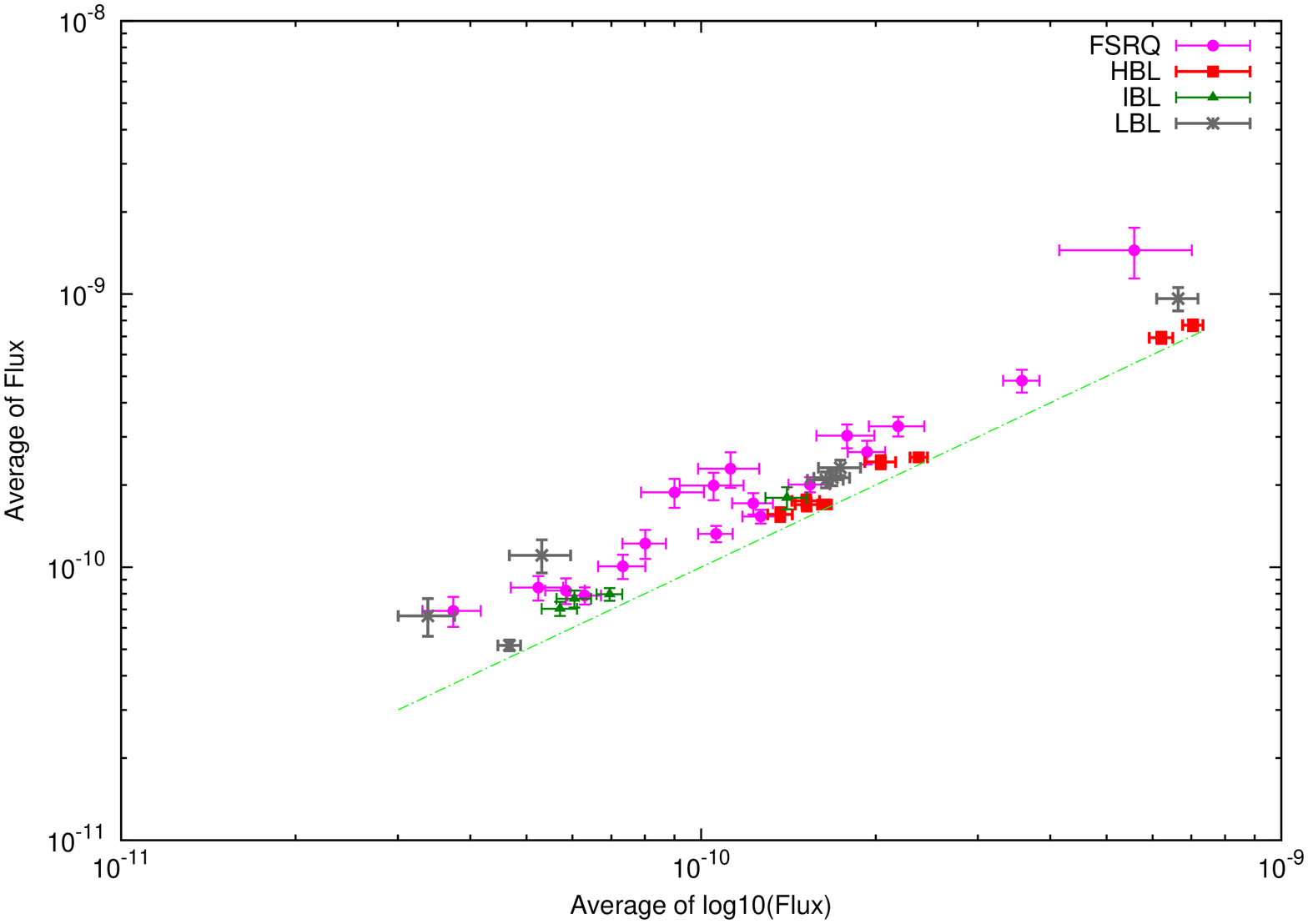}}\quad\hspace{-0.5cm}
	\subfigure{\includegraphics[scale=0.27,angle=270]{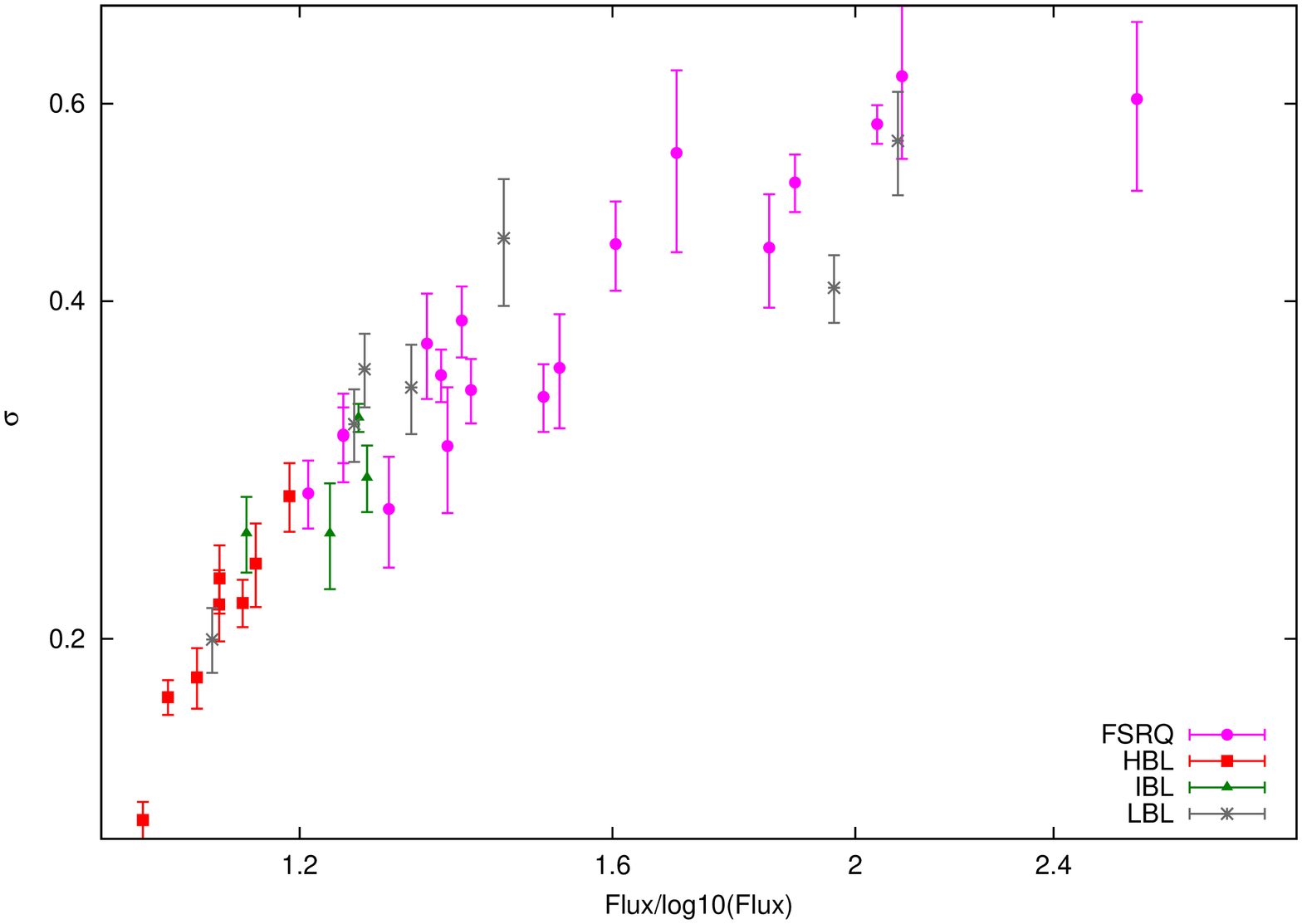}}  
	\caption{Left plot shows the average of flux vs average of logarithm of flux of bright blazars. The line shows the condition of equality of quantities. Standard deviation of the log-normal flux distribution is plotted against the ratio between flux and logarithm of flux. Pointer color indicates the blazar class, pink: FSRQ, red: HBL, green: IBL, and gray: LBL.} 
	\label{fig:flux_vs_flux}
\end{figure}

Another implication of our study is on the averaging of long term flux. We recommend the usage of the average of flux in log scale, rather than estimating average flux in a linear scale, especially for the highly variable sources. We show the difference of averaged flux in both linear and log-scale in Fig \ref{fig:flux_vs_flux}. For example, in the case of FSRQs, the average value of Flux/Log$_{10}$(Flux) falls around $\sim$1.7, while the maximum value (in the case of 3C\,454.3) goes up-to 2.8. These values imply that averaging flux over a linear scale will significantly overestimate the same, which would in turn gives rise to inaccurate SED non-thermal emission model parameters.

\section{Conclusion}

We studied in detail the flux distribution properties of 38 brightest $\gamma$-ray blazars using \emph{Fermi}-LAT data of more than 8 years. The flux distribution suggest log-normal distribution, for 35 blazars, indicating a multiplicative perturbation associated with the emission  process. Similar features were obtained also in the  case of BCUs. On the other hand, the flux distributions of three FSRQs -- J2329.3-4955, J1504.4+1029, and J1625.7-2527 -- reject both log-normal and normal distribution. This could be due to two or more independent flux states associated with the source,  however, more statistics is required to study these effects in detail. It would be also interesting to  perform an elaborate study with better statistics for more blazars in $\gamma$-rays, and compare the properties with that of their X-ray counterparts. 

\section{Acknowledgement}
ZS, SS and NI are thankful to Indian Space Research Organization program (ISRO-RESPOND) for the financial support under grant no. ISRO/RES/2/396.

\bibliographystyle{raa}
\bibliography{ms89}

\end{document}